\documentstyle[12pt]{article}

\begin{document}
\title{On the use of Mellin transform to a class of \\
q-difference-differential equations}
\author{Choon-Lin Ho\\
{\small \sl Department of Physics, Tamkang University, Tamsui 25137, Taiwan}}
\date{Mar 7, 2000}

\maketitle

\begin{abstract}
We explore the possibility of using the method of classical integral
transforms
to solve a class of $q$-difference-differential equations.  The Laplace and
the Mellin transform of $q$-derivatives are derived.  The  results  show that
the Mellin transform
of the $q$-derivative resembles most closely the corresponding
expression in classical analysis, and it could therefore be
useful in solving certain $q$-difference equations.
\end{abstract}
\vskip 2truecm
\leftline{Revised version}
\newpage

{\bf 1.}  The study of $q$-analysis is an old subject, which dates back to
the end of the 19th century (\cite{Rogers}-\cite{GR}).  It has found many
applications in
such areas as the theory of partitions, combinatorics, exactly solvable models
in statistical mechanics, computer algebra, etc \cite{Andrews}.  Recent
developments in the theory of quantum group has boosted further interests in
this old subject \cite{FV1,Ko}.

The subject of $q$-analysis concerns mainly the properties of the
so-called
$q$-special functions, which are the extensions of the classical special
functions based on a parameter, or the base, $q$.  The relations among these
functions,
and the difference equations satisfied by them are among the topics  most
studied so far.  The $q$-difference equations
involve a new kind of difference operator, the $q$-derivative,
which can be viewed as a sort of deformation of the ordinary derivative.
Solutions of the $q$-difference equations in one variable have been well
studied in terms of the $q$-hypergeometric series (also called the
basic hypergeometric series).  Partial $q$-difference equations and
$q$-difference-differential equations with more than
one variables are generally studied by means of the method of seperation
of variables, or by the techniques of Lie symmetry in
the literature (\cite{E},\cite{LVW}-\cite{BHNN}).   The method of integral
transforms, which is
another powerful technique of solving differential equations in
classical analysis, has not been, in our view, explored in $q$-analysis.
The reason is  not hard to understand.
The main virtue of the  classical integral transforms, particularly
the Fourier and the Laplace transform, is to transform a differential equation
into an algebraic equation, which can be solved easily.  That this is
possible is due to the fact that these transforms change the derivatives of
a function to something proportional to the transform of the original function.
As far as  we  know,
integral tranforms or $q$-integral transforms which could transform
$q$-difference equations into algebraic equations have not been found.
It should be mentioned that in fact $q$-analogues of Fourier transform,
based on the Jackson $q$-integral, have been proposed recently \cite{KS,K}.
However, in order for the
$q$-Fourier transform of the $q$-derivative of a function $f(x)$ to be
proportional to the $q$-Fourier transform of $f(x)$, the function $f(x)$
must satisfy
very special conditions, such as $f(q^{-1})=0=f(-q^{-1})$ \cite{K}.
Hence,  while these
$q$-Fourier transforms may be  useful in proving certain identities among the
$q$-special functions, their use in solving $q$-difference equations seems
limited.

In this paper we shall explore the possibility of using the method of
classical integral
transform to solve a class of $q$-difference-differential equations.
We derive the Laplace and the Mellin transform of $q$-derivative, and argue
that the Mellin transform, which is not generally employed in solving
differential equations in classical analysis, may still be useful in solving
certain $q$-difference equations.

\vskip 0.5  truecm

{\bf 2.}  Suppose we want to solve the following $q$-diffusion equation
\begin{eqnarray}
D^q_t y(x,t)=\frac{\partial^2}{\partial x^2}y(x,t)~~~(-\infty<x<\infty,~~t>0)
\label{Diffuse}
\end{eqnarray}
subject to the initial condition
\begin{eqnarray}
y(x,0)=f(x)~.
\end{eqnarray}
Here $D^q_t$ is the ``forward" temporal $q$-derivative defined by \cite{K,Koorn}
\begin{eqnarray}
D^q_t h(t):= \frac{h(q^{-1}t)-h(t)}{(1-q)t}~.
\label{qD-1}
\end{eqnarray}
for any function $h(x)$.
We assume $0<q<1$ in this paper.  The function $f(x)$ is assumed to vanish as
$x\to\pm\infty$.  One may as well use the more common
definition of $q$-derivative \cite{GR}
\begin{eqnarray}
{\cal D}^q_t h(t):= \frac{h(t)-h(qt)}{(1-q)t}~.
\label{qD-2}
\end{eqnarray}
We shall not employ this definition of the $q$-derivative here for reason to
be explained later.  We note here that $q$-difference and
$q$-difference-differential equations of the diffusion type such as
eq.(\ref{Diffuse}) have been considered before (\cite{LW}-\cite{BHNN}), but
mostly from the point of view of Lie symmetry, or by seperation of variables.

We can remove the partial differential operator in $x$ in (\ref{Diffuse})
by a Fourier transform.  The question now is to choose an appropriate integral
transform to remove the $q$-derivative.  In view  of the positivity of the
time variable, the two most natural choices are the Laplace and the Mellin
transform.

Let us first derive the expression of the Laplace
transform of the $q$-derivative.  The Laplace transform of a function $h(t)$
is defined as ${\bar h(s)}:={\cal L}\{h(x),s\}=\int_0^\infty h(t) \exp(-st)dt$.
For the $q$-derivative of $h(x)$, the Laplace transform is
\begin{eqnarray}
{\cal L}\{D^q_t h(t),s\}
=\frac{1}{1-q}\left[\int_0^\infty~\frac{h(q^{-1}t)}{t}e^{-st} dt
-\int_0^\infty~\frac{h(t)}{t}e^{-st} dt\right]~.
\label{L1}
\end{eqnarray}
To proceed we have to use the following relation of the Laplace transform
\cite{Sneddon}
\begin{eqnarray}
\int_s^\infty {\bar h}(s^\prime)ds^\prime=\int_0^\infty~\frac{h(t)}{t}e^{-st} dt~,
\label{L2}
\end{eqnarray}
provided the integral on the r.h.s. of (\ref{L2}) is well-defined.
We may apply (\ref{L2}) to (\ref{L1}) directly if $h(0)=0$.  However, if
$h(0)\neq 0$, the r.h.s. of (\ref{L2}) is not well-defined, and  direct
application of (\ref{L2}) to (\ref{L1})
leads to incorrect result which does not reduce to the usual expression
of the
Laplace transform of derivative in the classical limit $q\to 1^-$. In order to
recover the classical limit correctly, we find it necessary to regularise
(\ref{L1}) in the form
\begin{eqnarray}
\frac{1}{1-q}\left[\int_0^\infty~\frac{h(q^{-1}t)-h(0)}{t}e^{-st} dt
-\int_0^\infty~\frac{h(t)-h(0)}{t}e^{-st} dt\right]~.
\label{L3}
\end{eqnarray}
We may now apply (\ref{L2}) to (\ref{L3}).  Making use of
\begin{eqnarray}
{\cal L}\{h(t)-h(0),s\}={\bar h}(s)-s^{-1}h(0)
\end{eqnarray}
we finally obtained
\begin{eqnarray}
{\cal L}\{D^q_t h(t),s\}=\frac{1}{1-q}\int_{sq}^s~{\bar
h}(s^\prime)~ds^\prime - \frac{\ln q^{-1}}{1-q}h(0)~.
\label{L4}
\end{eqnarray}
Eq.(\ref{L4}) reduces to the expression $s{\bar h}(s)-h(0)$ for the Laplace
transform of ordinary derivative as $q\to 1^-$.

If one uses instead the definition (\ref{qD-2}) for the $q$-derivative, the
Laplace transform would be
\begin{eqnarray}
{\cal L}\{{\cal D}^q_t h(t),s\}=\frac{1}{1-q}\int_s^{\frac{s}{q}}~{\bar
h}(s^\prime)~ds^\prime - \frac{\ln q^{-1}}{1-q}h(0)~.
\end{eqnarray}

It is now obvious that the Laplace transform is not useful in solving
equations involving $q$-derivatives: it transforms such equations into
integral equations!

\vskip 0.5  truecm

{\bf 3.}  We now consider the Mellin transform of a $q$-derivative. The
Mellin transform is seldom being used in solving differential equations,
because it generally transforms differential equations into difference
equations instead of the much simpler algebraic equations.
Now that the Fourier and the Laplace transform lose their virtues whenever
$q$-derivatives are present, the Mellin transform is naturally the next
one to be looked at.  As we shall see below, the Mellin transform still
transforms an equation containing $q$-derivatives into a difference equation
of the transformed function, which is the best thing next to an algebraic
equation one could get.  Previously, the use of the Mellin transform in
$q$-analysis is
limited to proving various identities among the $q$-special functions
\cite{S,AAR}.

The Mellin transform
of a function $h(t)$ is defined as $h^*(s):={\cal M}\{h(t),s\}=\int_0^\infty
h(t) t^{s-1}dt$. For $q$-derivative defined in  (\ref{qD-1}), we have
\begin{eqnarray}
{\cal M}\{D^q_t h(t),s\}=-[s-1]_q~h^*(s-1)~.
\label{Mellin}
\end{eqnarray}
Here $[x]_q$ is the $q$-number defined by
\begin{eqnarray}
[x]_q:=\frac{1-q^x}{1-q}~.
\end{eqnarray}
Note that $[x]_q\to x$ as $q\to 1^-$.
Hence (\ref{Mellin}) reduces to the expression $-(s-1)h^*(s-1)$ for the Mellin
transform of the ordinary derivative as $q\to 1^-$.  Repeated use of
(\ref{Mellin}) leads to
\begin{eqnarray}
{\cal M}\{(D^q_t)^n h(t),s\}=(-1)^n[s-1]_q[s-2]_q\cdots[s-n]_q~h^*(s-n)~,
~~~n\geq 1~.
\end{eqnarray}
This is the $q$-analogue of  the corresponding formula in the classical
case \cite{Sneddon}.

For the definition (\ref{qD-2}),  one has
\begin{eqnarray}
{\cal M}\{{\cal D}^q_t h(t),s\}&=&[1-s]_q~h^*(s-1)\\
&=&-q^{1-s}[s-1]_q~h^*(s-1)~.
\end{eqnarray}
Here an extra factor of $q$ appears compared with (\ref{Mellin}).  In order
to simplify our presentation, we therefore adopt the definition (\ref{qD-1})
in this paper.    We must, however, mention that all the arguments
given below apply equally well to the corresponding cases with
$q$-derivatives replaced by the definition (\ref{qD-2}).

\vskip 0.5  truecm

{\bf 4.} Let $Y^* (\xi, s)$ be the transformed function of $y(x,t)$
obtained by taking the Mellin transform in $t$ and a Fourier transform
$G(\xi):=\int^\infty_{-\infty} g(x)\exp(i\xi  x)dx$ in $x$. Making these
transforms to (\ref{Diffuse}),  one obtains
\begin{eqnarray}
[s-1]_q Y^*(\xi,s-1)=\xi^2 Y^*(\xi,s)~.
\end{eqnarray}
Fortunately solution to this equation can be readily found to be
\begin{eqnarray}
Y^*(\xi,s)=A(\xi)\xi^{-2s}\Gamma_q (s)~,
\label{M-Y}
\end{eqnarray}
where $A(\xi)$ is some function of $\xi$ only, and $\Gamma_q (s)$ is the
{\sl q-gamma function} defined by \cite{GR}
\begin{eqnarray}
\Gamma_q (s):=\frac{(q;q)_\infty}{(q^s;q)_\infty}\left(1-q\right)^{1-s}~~,~~0<q<1~.
\end{eqnarray}
\begin{eqnarray}
(a;q)_\infty:=\prod_{k=0}^\infty(1-aq^k)~.
\end{eqnarray}
$\Gamma_q(s)$ satisfies
\begin{eqnarray}
\lim_{q\to 1^-} \Gamma_q (s) &=& \Gamma (s)~,\\
\Gamma_q (s+1)&=&  [s]_q \Gamma(s)~,~~~\Gamma_q (1)=1~.
\end{eqnarray}
Inverse-Mellin transform of $\xi^{-2s} \Gamma_q(s)$ in (\ref{M-Y}) is
\begin{eqnarray}
\frac{1}{2\pi i}\int_{-i\infty}^{i\infty} \xi^{-2s}\Gamma_q(s) t^{-s}ds~.
\label{I-Mellin}
\end{eqnarray}
The poles of $\Gamma_q(s)$ are $s=0,-1,-2,\ldots$. The residual of
$\Gamma_q(s)$ at pole $s=-n$ ($n\geq 0$) is \cite{GR}:
\begin{eqnarray}
\frac{(1-q)^{n+1}}{(q^{-n};q)_n \ln q^{-1}}~.
\end{eqnarray}
The symbol $(a;q)_n$ is the {\sl q-shifted factorial}:
\begin{eqnarray}
&&(a;q)_0:=1~,~~~~~~~~~n=0~,\\
(a;q)_n&:=& (1-a)(1-aq)\cdots(1-aq^{n-1})~,~~ n=1,2\ldots
\end{eqnarray}
Hence (\ref{I-Mellin}) becomes
\begin{eqnarray}
\frac{1-q}{\ln q^{-1}}~\sum_{n=0}^\infty \frac{\left[(1-q)\xi^2 t\right]^n}
{(q^{-n};q)_n}~.
\label{sum1}
\end{eqnarray}
In view of the identity \cite{GR}
\begin{eqnarray}
(q^{-n};q)_n= \left(-\frac{1}{q}\right)^nq^{-n(n-1)/2} (q;q)_n~,
\end{eqnarray}
(\ref{sum1}) can be expressed as
\begin{eqnarray}
&&\frac{1-q}{\ln q^{-1}}~\sum_{n=0}^\infty \frac{q^{n(n-1)/2}}
{(q;q)_n}\frac{\left[-q(1-q)\xi^2 t\right]^n} \nonumber\\
&=&\frac{1-q}{\ln q^{-1}}~E_q\left(-q(1-q)\xi^2 t\right)~.
\label{sum2}
\end{eqnarray}
The function $E_q(z)$ (for complex $z$) is the {\sl q-exponential function}
defined by \cite{GR}
\begin{eqnarray}
E_q(z):=\sum_{n=0}^\infty \frac{q^{n(n-1)/2}z^n}{(q;q)_n}=(-z;q)_\infty~.
\label{q-Exp}
\end{eqnarray}
In the limit $q\to 1^-$, eq.(\ref{sum2}) tends to
the usual exponential function $\exp(-\xi^2 t)$.  Finally, performing an
inverse Fourier transform we obtain the solution of the $q$-diffusion equation
\begin{eqnarray}
y(x,t)=\frac{1}{\sqrt{2\pi}}\int_{-\infty}^{\infty} A(\xi)
\left\{\frac{1-q}{\ln q^{-1}}~E_q\left(-q(1-q)\xi^2 x\right)\right\}
e^{-i\xi x}~d\xi~.
\label{Y1}
\end{eqnarray}
Setting $t=0$ in (\ref{Y1}) shows that
\begin{eqnarray}
\frac{1-q}{\ln q^{-1}}~A(\xi)&=&\frac{1}{\sqrt{2\pi}}\int_{-\infty}^{\infty}
y(x,0)e^{i\xi x}~dx\nonumber\\
&=&\frac{1}{\sqrt{2\pi}}\int_{-\infty}^{\infty}
f(x) e^{i\xi x}~dx\nonumber\\
&\equiv& F(\xi)
\end{eqnarray}
is the Fourier transform of $y(x,0)=f(x)$.
So the final solution of the initial problem is
\begin{eqnarray}
y(x,t)=\frac{1}{\sqrt{2\pi}}\int_{-\infty}^{\infty} F(\xi)
E_q\left(-q(1-q)\xi^2 t\right)
e^{-i\xi x}~d\xi~.
\label{Y2}
\end{eqnarray}
This is the $q$-analogue of the solution given in \cite{Titch} for the
corresponding classical case.
One can easily check that (\ref{Y2}) indeed satisfies (\ref{Diffuse}) by using
the following identity
\begin{eqnarray}
D^q_t E_q (\lambda t)=\frac{\lambda}{q(1-q)} E_q(\lambda t)~.
\label{D-E}
\end{eqnarray}

Let us consider an example. Suppose the initial profile is
$f(x)=\exp(-x^2/4b)/\sqrt{2b}$, ($b>0$).  Its Fourier transform is
$F(\xi)=\exp(-b\xi^2)$.  Then from (\ref{Y2}) and (\ref{q-Exp}), we get
\begin{eqnarray}
y(x,t)=E_q\left(q(1-q)t\frac{d}{db}\right)f(x)~.
\label{Y3}
\end{eqnarray}
In the limit $q\to 1^-$, eq.(\ref{Y3}) gives the classical solution
\begin{eqnarray}
y(x,t)&=&e^{t\frac{d}{db}}\left(\frac{1}{\sqrt{2b}}e^{-\frac{x^2}{4b}}\right)
\nonumber\\
&=& \frac{1}{\sqrt{2(t+b)}}e^{-\frac{x^2}{4(t+b)}}~.
\end{eqnarray}

\vskip 0.5  truecm

{\bf 5.} As another example, let us consider the following wave equation
\begin{eqnarray}
\left(D^q_t\right)^2 y(x,t)=\frac{\partial^2}{\partial x^2}~y(x,t)
~~~(-\infty < x <\infty~,~t>0)
\label{wave}
\end{eqnarray}
with inital conditions
\begin{eqnarray}
y(x,0)=f(x),~~~~D^q_t y(x,0)=g(x)~.
\end{eqnarray}
We assume that both $f(x)$ and $g(x)$ vanish as $x\to \pm\infty$.
In this case the Fourier-Mellin transformed function $Y^* (\xi,s)$ obeys
\begin{eqnarray}
[s-1]_q [s-2]_q Y^*(\xi,s-2)=-\xi^2 Y^*(\xi,s)~.
\end{eqnarray}
The general  solution is
\begin{eqnarray}
Y^*(\xi,s)=\left[A(\xi)\left(-i\xi\right)^{-s}+B(\xi)\left(i\xi\right)^{-s}
\right]~\Gamma_q (s)~,
\end{eqnarray}
where $A(\xi)$ and $B(\xi)$ are some functions of $\xi$.
Performing the inverse-Mellin transform, we get
\begin{eqnarray}
Y(\xi,t)=\frac{1-q}{\ln q^{-1}}\Bigl\{A(\xi)E_q\left(iq(1-q)\xi t\right)+
B(\xi)E_q\left(-iq(1-q)\xi t\right)\Bigr\}~.
\label{Y4}
\end{eqnarray}
Here $Y(\xi,t)$ is the Fourier transform of $y(x,t)$ with respect to $x$.
Now we rewrite (\ref{Y4}) in terms of the {\sl q-Sine} and
the {\sl q-Cosine} function which are defined by \cite{GR}
\begin{eqnarray}
{\rm~Sin}_q (x)=\frac{E_q(ix)-E_q(-ix)}{2i}~,\\
{\rm~Cos}_q (x)=\frac{E_q(ix)+E_q(-ix)}{2}~.
\end{eqnarray}
The result is
\begin{eqnarray}
y(\xi,t)=\frac{1-q}{\ln q^{-1}}\Bigl\{C(\xi){\rm~Cos}_q\left(q(1-q)\xi
t\right)+ D(\xi){\rm~Sin}_q\left(q(1-q)\xi t\right)\Bigr\}~,
\label{Y5}
\end{eqnarray}
where the functions $C(\xi)$  and $D(\xi)$ are linear combinations of $A(\xi)$
and $B(\xi)$.  The inverse-Fourier transform of (\ref{Y5}) is
\begin{eqnarray}
y(x,t)=\frac{1-q}{\sqrt{2\pi}\ln q^{-1}}
\int_{-\infty}^\infty\Bigl\{C(\xi){\rm~Cos}_q\left(q(1-q)\xi
t\right)+ D(\xi){\rm~Sin}_q\left(q(1-q)\xi t\right)\Bigr\}~
e^{-i\xi x}~d\xi~.
\label{Y6}
\end{eqnarray}
Letting $t=0$ in (\ref{Y6}), one can check that the function $C(\xi)$ is
related to the Fourier transform of $f(x)$ by
\begin{eqnarray}
F(\xi)=\frac{1-q}{\ln q^{-1}}~C(\xi)~.
\end{eqnarray}
Making use of the following relations,
which can be obtained by means of (\ref{D-E}):
\begin{eqnarray}
D^q_t {\rm~Sin}_q (\lambda t) &=& \frac{\lambda}{q(1-q)}{\rm~Cos}_q(\lambda
t)~,\\
D^q_t {\rm~Cos}_q (\lambda t) &=&-\frac{\lambda}{q(1-q)}{\rm~Sin}_q(\lambda
t)~,
\end{eqnarray}
we can relate $D(\xi)$ to the Fourier transform $G(\xi)$ of $g(x)$ as follows:
\begin{eqnarray}
G(\xi)=\frac{1-q}{\ln q^{-1}}~D(\xi)\xi~.
\end{eqnarray}
With these results, we finally obtain the solution to the initial problem of
eq.(\ref{wave}):
\begin{eqnarray}
y(x,t)=\frac{1}{\sqrt{2\pi}}
\int_{-\infty}^\infty\left\{F(\xi){\rm~Cos}_q\left(q(1-q)\xi
t\right)+ \frac{G(\xi)}{\xi}{\rm~Sin}_q\left(q(1-q)\xi t\right)\right\}~
e^{-i\xi x}~d\xi~.
\label{Y7}
\end{eqnarray}
This solution is the $q$-analogue of the solution to the corresponding
classical case given in \cite{Titch}.

\vskip 0.5  truecm

{\bf 6.} We now see how the above steps are generalised to the equation:
\begin{eqnarray}
\left(D^q_t\right)^n y(x,t)=\frac{\partial^2}{\partial x^2}~y(x,t)
~~~(-\infty < x <\infty~,~t>0,~~ n\geq 2)
\end{eqnarray}
with inital conditions
\begin{eqnarray}
y(x,0)=f(x),~~~~\left(D^q_t\right)^k y(x,0)=g_k (x)~,~~~k=1,\ldots,n-1~,
\end{eqnarray}
where the functions $f(x)$ and $g_k(x)$ are assumed to vanish as $x\to
\pm\infty$.
The Fourier-Mellin transformed function $Y^* (\xi,s)$ obeys
\begin{eqnarray}
(-1)^n[s-1]_q[s-2]_q\cdots[s-n]_q Y^*(\xi,s-n)=-\xi^2 Y^*(\xi,s)~.
\end{eqnarray}
The general  solution is
\begin{eqnarray}
Y^*(\xi,s)=\Gamma_q (s)\xi^{-\frac{2s}{n}}~
\sum_{m=0}^{n-1}~A_m(\xi)\left[-e^{-\frac{(2m+1)}{n}\pi i}\right]^s~.
\end{eqnarray}
where $A_m(\xi)$ are some functions of $\xi$.
We can now perform the inverse Mellin and Fourier transforms to get the final
solution, which is given formally as
\begin{eqnarray}
y(x,t)=\frac{1-q}{\sqrt{2\pi}\ln q^{-1}}\sum_{m=0}^{n-1}\int_{-\infty}^{\infty}
A_m(\xi) E_q\left(q(1-q)e^{\frac{(2m+1)}{n}\pi i}\xi^{\frac{2}{n}}t\right)
e^{-i\xi x}~d\xi~.
\end{eqnarray}
The functions $A_m(\xi)$ can then be related to the Fourier
transforms of the functions $f(x)$ and $g_k(x)$ from the initial conditions.

\vskip 0.5  truecm

{\bf 7.}  To summarise, we show that the Mellin
transform of the $q$-derivative resembles most closely the corresponding
expression in classical analysis, whereas transforms such as the Fourier
and  the  Laplace transform fail in this respect.  As such the Mellin
transform can be useful in solving certain $q$-difference equations. We
illustrated this fact with a few examples.  However, for the Mellin transform
to be really useful, a more complete knowledge of the properties of
the $q$-special functions under various integral transforms (Fourier, Laplace,
Mellin, etc) and  their inverses has yet to be attained.  What is more
desirable is to invent integral transforms or $q$-integral transforms that
possess the virtue of the  Fourier and the Laplace transform in the classical
analysis mentioned in the introduction.

\vskip 2 truecm
\centerline{\bf Acknowledgment}

This work is supported in part by the Republic of China through Grant
No. NSC-89-2112-M-032-004.

\end{document}